\documentclass{WileyMSP-template}

\usepackage{amsmath}
\usepackage{multirow}
\usepackage{array}
\usepackage{graphicx}
\usepackage{booktabs}
\usepackage{caption}
\usepackage{comment}

\newcommand{\mum}{\mu\mathrm{m}}

\begin{document}

\pagestyle{fancy}
\rhead{\includegraphics[width=2.5cm]{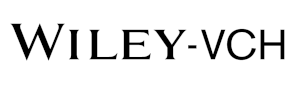}}

\title{Optimizing Phononic Crystal Waveguides for Acoustically Induced Spin Transport}

\maketitle


\author{Karanpreet Singh}
\author{Gabe Wilson}
\author{James A.H. Stotz*}


\dedication{}

\begin{affiliations}
K.~Singh, G.~Wilson, J.~A.~H.~Stotz\\
Department of Physics, Engineering Physics \& Astronomy, Queen's University \\ Kingston, ON  K7L 3N6, Canada \\
jstotz@queensu.ca


\end{affiliations}


\keywords{phononic crystal, waveguide, finite element method, spin transport}


\begin{abstract}

Through the use of strain and induced piezoelectric fields, surface acoustic waves have been shown to control quantum information processes, such as single photon emission and the coherent transport of electron spins.  Regarding the latter, systems using plane surface waves have provided suitable demonstration systems, but to build complexity, more control over the acoustic wave may be required.  One method for acoustic control is the use of phononic crystals consisting of periodic arrays of nanofabricated holes on the surface of a device. These inclusions form a metamaterial-like layer with properties different from the normal material to dictate the physics of wave motion. Exploiting these surface properties can lead to acoustic waveguides, which can be designed to control the path of the surface acoustic waves.  The design parameters of a new type of phononic crystal waveguide is explored that uses 2-fold elliptical cylinder inclusions to create a slow region that also limits coupling and radiative loss to bulk acoustic modes.  Such a waveguide would be the foundational piece in an acoustic circuit that could then mediate complex spin transport geometries.

\end{abstract}

\section{Introduction}

The desire for scalability in quantum computation is illustrated by the fact that an estimated 2~million qubits are needed for computations in quantum chemistry \cite{Beverland2022}.  To address this issue, groups have been developing ways to move quantum information around devices using ions\cite{Akhtar2023}, electrons\cite{Wang2023}, and spins\cite{Helgers22}. The latter two studies use acoustic excitations in the form of surface acoustic waves (SAWs) to confine and transport quantum states.  SAWs are effective as their strain and piezoelectric fields can couple to semiconductor nanostructures and the charges and spins they contain.  The high frequencies of the SAWs enables them to transport on the order of 1~billion individual electrons per second, therefore providing an interesting paradigm to study for quantum computation \cite{Barnes00}.

\ 

To this end, this ability of SAWs to couple to quantum states in semiconductors has lead to a number of interesting applications such as: the creation of single electron transistors\cite{Shilton1996}; control of single photons from quantum dots\cite{Gell2008, Couto09}; opening additional band gaps in photonic structures\cite{deLima2005} and inducing the formation of a Bose-Einstein condensate\cite{CerdaMendez2010}. In those studies using acoustically mediated charge transport, the electrons and holes are spatially separated by the piezoelectric electric field induced by the SAW and do not interact, allowing the ambipolar transport to extend up to millimeters due to the significant increase in their recombination lifetimes\cite{Rocke1997}. When a lateral confinement is applied (transverse to the direction of motion), the charges are then transported in fully 3-D confinement potentials, referred to as dynamic quantum dots (DQDs), that form a quantum conveyor belt and can be used to transport single electrons\cite{Shilton1996, Hermelin11, McNeil2011} or to coherently transport electron spins \cite{Helgers22, Stotz05}. However, previous work involving electron spins involved only simple, linear transport. To create non-trivial paths for the flying (as opposed to stationary) electron spins, a system for acoustic control could be based on controlling SAWs within phononic crystal (PnC) waveguide structures \cite{Benchabane15}. By creating acoustic circuits compatible with semiconductor nanostructures, complex, acoustically mediated circuits could be realized to control the flow of spin information.

\ 

The proper design of a PnC structures is key as they cannot adversely affect the semiconductor nanostructure in an effort to control the SAW.  For example, the transport of electron spins uses a GaAs quantum well as the transport layer, and a phononic crystal inclusion would require the use of shallow inclusions \cite{Petrus14} that would not puncture the quantum well or resonant pillars on the surface \cite{AlLethawe16}.  Naturally, the confined acoustic mode must also generate a piezoelectric field in the region below the surface where the quantum well exists.  Using shallow inclusions, a waveguide that simply consisted of the PnC metamaterial could confine a SAW and generate the requisite piezoelectric fields below the surface \cite{Muzar23}.  While being able to waveguide, that study still suffered from energy loss due to acoustic coupling to bulk modes. Recently, a new 2-fold symmetric inclusion geometry was proposed that lowers the mode frequencies below the sound line thus dramatically reducing energy loss while demonstrating an improved waveguide confinement \cite{Singh2024}.  Using this new inclusion design, the design and optimization of a waveguide will be examined with specific consideration made for the applications of acoustic circuits mediating spin transport in GaAs.

\ 

In the original work using shallow inclusions as the phononic crystal waveguide in GaAs \cite{Muzar23}, it was found that a waveguide of four unit cells wide provide the best balance of lateral confinement while maintaining a core strain energy ratio above 0.6 and have the highest reciprocal attenuation. The original report of using elliptical cylinder inclusions replicated the four inclusion waveguide width for comparison \cite{Singh2024}, but a more detailed investigation was not completed. Building on that work, here we will explore the impact of waveguide parameters such as waveguide width, inclusion aspect ratio, and inclusion depth on key metrics characterizing surface acoustic wave (SAW) confinement, dispersion, and bandwidth within the waveguide, aiming to optimize waveguide performance for enhanced acoustic spin transport in GaAs.

\section{Waveguide Width Optimization}

\begin{figure}[bt]
    \centering
    \includegraphics[width=0.5\linewidth]{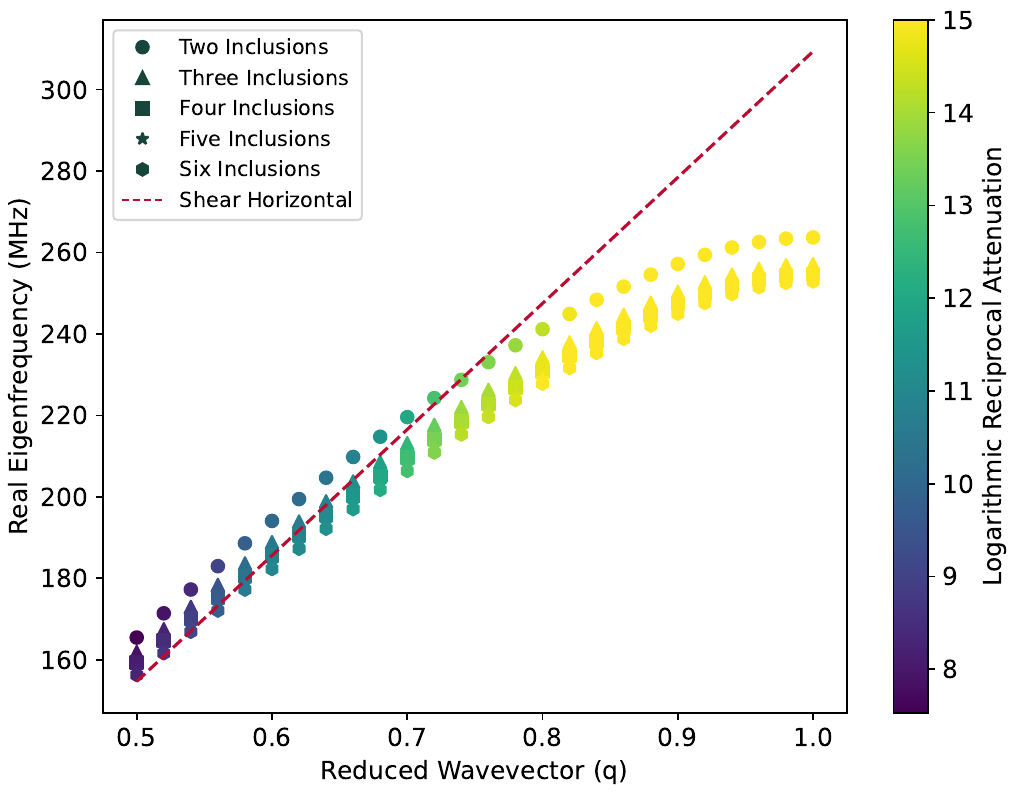}
    \caption{Dispersion relations showing real eigenfrequencies of the Rayleigh-like SAW modes versus reduced wavevector ($q$) for waveguides with widths of two to six inclusions. Each unit cell of the waveguide includes an elliptical cylinder inclusion (semi-minor axis of $0.566~\mum$ and a semi-major axis of $1.7~\mum$) that is $3~\mum$ deep. Colour scale indicates logarithmic reciprocal attenuation.}
    \label{dispersion-relations}
\end{figure}

The width of the waveguide is defined by adjacently placed unit cells each containing an identical inclusion. A lattice parameter of $a = 4~\mum$ is being used forming waveguides with widths between $8~\mum$ to $24~\mum$ for structures having two to six inclusions, respectively.  Basic dispersion relations were calculated for these waveguide widths to determine the lowest order SAW modes in these structures.  Figure \ref{dispersion-relations} shows the real eigenfrequencies of the Rayleigh-like SAW modes for elliptical cylinders with an aspect ratio of $3:1$ as a function of the reduced wavevector $q$ ($q = k/k_x$, where $k$ is the wavevector and $k_x = \pi/a$). The dispersion diagram spans $q$ values from $0.5$ to $1.0$, with the colour scale indicating logarithmic reciprocal attenuation - the figure of merit that is used to quantify energy loss to the bulk. Higher reciprocal attenuation indicates stronger vertical SAW confinement at the surface with minimal coupling to bulk acoustic modes.

\ 


As the number of inclusions comprising the waveguides increases, Fig.~\ref{dispersion-relations} shows that eigenvalues are becoming smaller and converging to a particular value.  This would suggest that the wider waveguides contain a larger proportion of the SAW as the SAW inherits the properties of the waveguide (including the lower eigenfrequencies) and less of the bare substrate.  This is most notable for the two- and three-inclusion wide waveguides, while the remaining waveguides are quite similar.  

\

This lateral confinement also impacts the ability of the waveguide to confine the SAW to the surface.  The shear horizontal threshold, shown by the red line in Fig.~\ref{dispersion-relations}, is the slowest bulk mode to which the SAW may couple.  As the SAW modes transition below this line at higher $q$, there is a significant decrease in the reciprocal attenuation indicating minimal energy loss and enhanced surface confinement for these modes. For two- and three-inclusion wide waveguides, eigenfrequencies fall below the SH mode threshold only at higher $q$ values (above $0.76$ and $0.7$, respectively). In contrast, waveguides with four or more inclusions demonstrate this behavior at lower wavelengths, starting around $q = 0.64$ and almost remaining at this value even with increasing waveguide width. The logarithmic reciprocal attenuation is consistently higher at higher wavelengths in wider waveguides (four to six inclusions) compared to narrower two- and three-inclusion structures. This indicates that wider waveguides more effectively slow down the SAWs below the SH threshold, thus improving vertical confinement.

\

\begin{figure}[tb]
    \centering
    \includegraphics[width=0.5\linewidth]{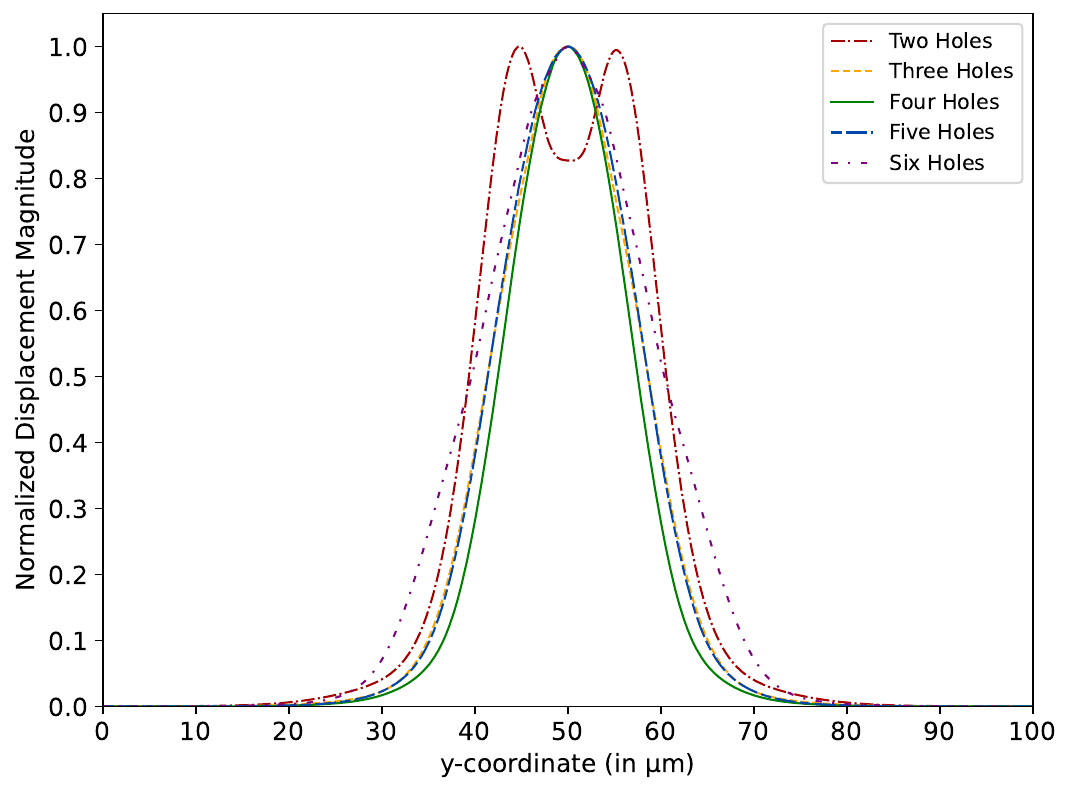}
    \caption{Normalized displacement magnitude profiles across waveguide width for two to six inclusion configurations, centered at $y = 50 \mum$. The profiles correspond to the $x=2~\mum$ face of the computational domain and at $z=-5.4~\mum$, the latter of which corresponds to half the wavelength at $q=0.74$ }
    \label{line-graph}
\end{figure}

While wider waveguides demonstrate improved vertical confinement by minimizing leakage into the bulk substrate, effective lateral confinement of the SAW within the waveguide core is crucial for electron spin transport \cite{Helgers22, Stotz08}. Precise control over the spatial distribution of the SAW enables a narrow pathway for spin-polarized carriers, minimizing intrinsic D’yakonov–Perel’ spin dephasing thus increasing spin coherence lifetimes \cite{Dyakonov1972}.  To therefore assess the lateral confinement characteristics of the waveguides, the normalized displacement magnitude profiles across the waveguide width ($y$-coordinate) are analyzed with waveguide cores centered at $y = 50 \mum$, as shown in Fig.~\ref{line-graph}.  The profiles correspond to the $x=2~\mum$ of the computational domain (i.e., at the centre of the domain) and at $z=-5.4~\mum$, the latter of which corresponds to half the wavelength at $q=0.74$.  Taking the profiles at this position removes local minima or maxima resulting from the inclusion although the profile may be slightly wider than at the surface.

\


The four-inclusion waveguide exhibits the strongest lateral confinement, with the SAW energy mostly concentrated within its $16~\mum$ wide core. The three- and five-inclusion waveguides show similar confinement characteristics, although stemming from different underlying causes. While the five-inclusion waveguide confines the SAW within its larger $20~\mum$ core, the three-inclusion waveguide does have some strong confinement but the velocity differential with respect to the bare surface is not sufficient, and acoustic energy leaks outside its $12~\mum$ core. Interestingly, the two-inclusion waveguide displays a double-peaked profile, with two centers of high displacement located outside the core region, flanking the inclusions. In contrast, the six-inclusion waveguide shows a decrease in the slope profile allowing the mode to spread laterally, even though it is fully contained in the wide waveguide. Here, the width of the waveguide is over twice as large as the wavelength of the eigenmode, and waveguide is no longer restricting the wave as effectively to only the propagation direction.

\ 

\


\begin{figure}[bt]
    \centering
    \includegraphics[width=0.5\linewidth]{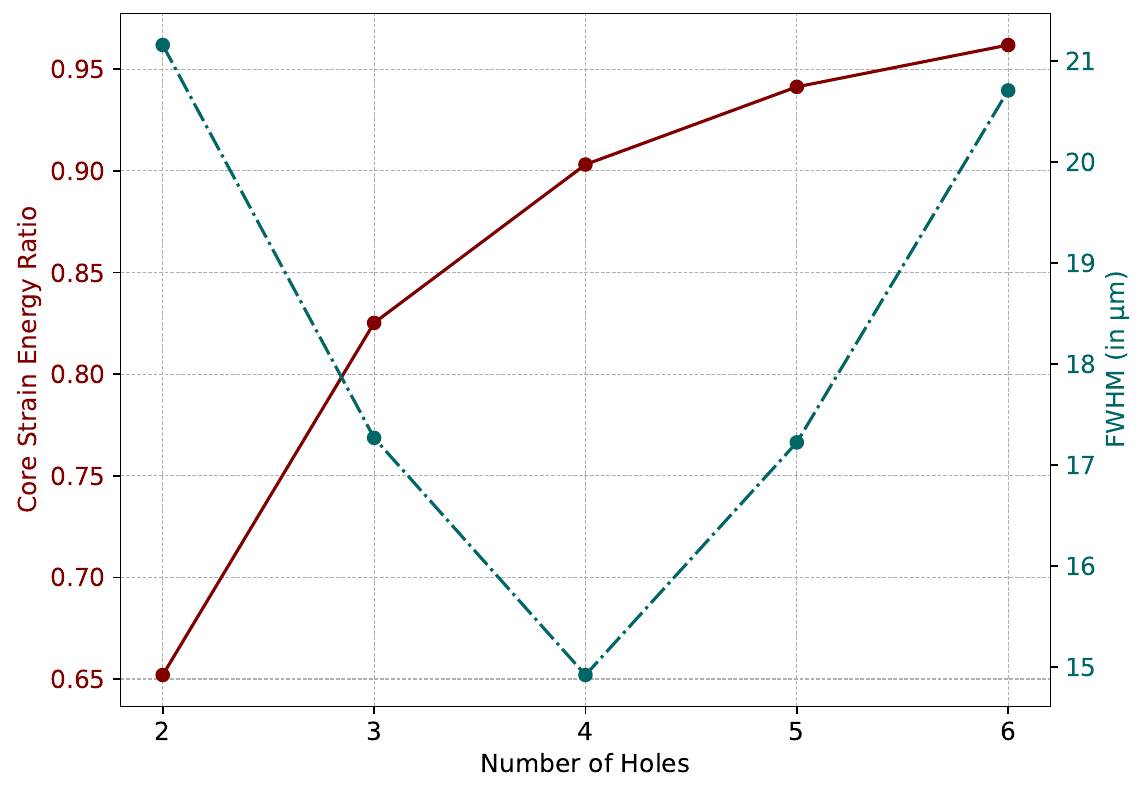}
    \caption{Core strain energy ratio and full width at half maximum (FWHM) as a function of number of inclusions in the waveguide. The FWHM values are derived from the profiles of Fig.~\ref{line-graph}.}
    \label{cser-fwhm}
\end{figure}


To further quantify the observations, the core strain energy ratio and the full width at half maximum (FWHM) of the displacement magnitude profile are determined as a function of the number of inclusions, as shown in Figure \ref{cser-fwhm}. The core strain energy ratio represents the fraction of the total strain energy confined within the waveguide core in the upper $25~\mum$ of the computational domain. As the number of inclusions increases, the core strain energy ratio also increases, indicating that wider waveguides effectively reduce energy leakage into the cladding and - to a lesser extent - into the bulk, consistent with the observations from Figs.~\ref{dispersion-relations}~and~\ref{line-graph}.  Interestingly, the two- and three-inclusion wide waveguides studied for cylindrical inclusion were not considered in previous work as they did not achieve the 60\% threshold for the core strain energy ratio \cite{Muzar23}.  This contrasts with the current case where both exceed that threshold.  

\ 

While it is useful to know how much of the wave is within the designed waveguide region, the core strain energy ratio is a somewhat imperfect as it is a relative measure.  As the waveguide becomes larger, there is a larger region for the acoustic energy to occupy. As a complimentary parameter to the core strain energy ratio, the FWHM provides a quantitative measure of the lateral confinement of the SAW within the waveguide core, and it is calculated from the normalized displacement magnitude profiles of Fig.~\ref{line-graph}. A lower FWHM signifies an tighter lateral confinement, as the wave energy is physically more concentrated within a narrower region of the waveguide core. Notably, the four-inclusion waveguide exhibits the lowest FWHM, confirming its superior lateral confinement characteristics as observed in Fig.~\ref{line-graph}; it is also the narrowest waveguide where the FWHM is less than the width of the waveguide core. The combination of high core energy confinement and low FWHM, therefore, demonstrates that a $16~\mum$ four-inclusion wide waveguide offers the best balance between vertical and lateral confinement for efficient SAW guiding. 

\

\begin{figure}[tb]
    \centering
    \includegraphics[width=0.6\linewidth]{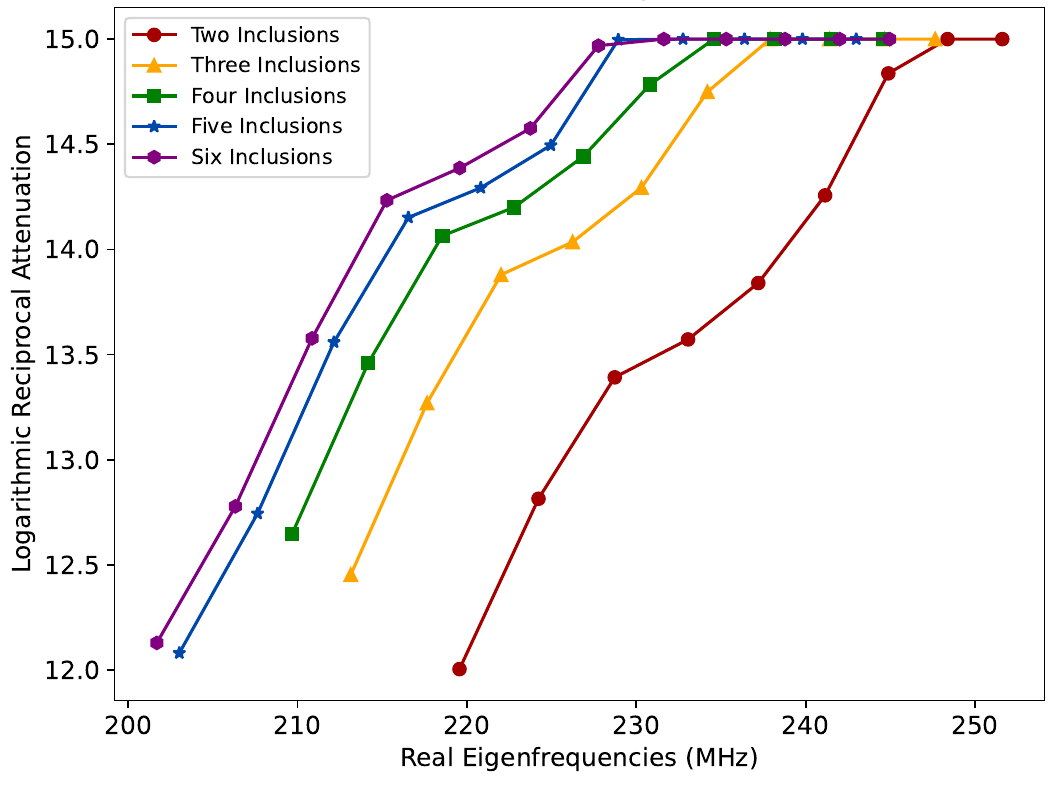}
    \caption{Operational bandwidth analysis for varying waveguide widths (two to six inclusions). Only frequencies from eigenmodes that meet both vertical confinement (log reciprocal attenuation $> 12$) and group velocity criteria ($v_g > \frac{v_p}{2}$ are included.}
    \label{bandwidth-diagram-inclusion-width}
\end{figure}

Rather than discussing the operation of these waveguides relative to their wavevectors $k$ or $q$, it would be more practical to discuss the frequency range for which the waveguides may operate.  To determine the optimal operating range for each waveguide width, we examine the bandwidth - defined as the range of frequencies over which a waveguide effectively confines and guides SAWs with minimal radiative loss to bulk modes.  Figure \ref{bandwidth-diagram-inclusion-width} illustrates the relationship between eigenfrequency and logarithmic reciprocal attenuation for each waveguide configuration. In this analysis, we impose two key criteria to ensure both strong vertical confinement and efficient information transport. First, only eigenfrequencies are accepted when the logarithmic reciprocal attenuation exceeds 12, which ensures that propagating SAWs are strongly confined to the surface.  As can be seen in Fig.~\ref{dispersion-relations} and in previous work \cite{Singh2024}, this is the value of the logarithmic reciprocal attenuation that is achieved once the eigenfrequencies fall below that of the shear-horizontal mode.  On the upper end of the frequency range, which also corresponds to large wavevectors, we approach the region near the edge of the Brillouin zone, which should be avoided.  Operating near the Brillouin zone edge is undesirable not only because the SAW group velocity approaches zero at the edge but also due to increased dispersion effects which can distort the SAW signal. As such, the upper edge of the bandwidth will be restricted to when the group velocity is less than half the phase velocity of the propagating wave. 

\


As evident from Figure \ref{bandwidth-diagram-inclusion-width}, wider waveguides exhibit a wider bandwidth. This increase in bandwidth is attributed to the waveguides reaching much lower frequencies, a direct consequence of the reduced eigenfrequencies in the dispersion relation (Fig.~\ref{dispersion-relations}), which presents a broad range of frequencies below the shear horizontal mode. Furthermore, each waveguide width exhibits a distinct operational frequency range, although with some overlap between structures. Table \ref{bandwidth-table-inclusion-width} summarizes the minimum and maximum eigenfrequencies and the corresponding bandwidth for each waveguide width. 

\


\begin{table}[tb]
 \caption{Operational Bandwidth for Waveguides With a Varying Number of Inclusions}
    \begin{center}
    \begin{tabular}{|c|c|c|c|}
        \hline
        \textbf{Number of Inclusions} & \textbf{Minimum Frequency (MHz)} & \textbf{Maximum Frequency (MHz)} & \textbf{Bandwidth (MHz)} \\
        \hline
        Two & 219.56 & 251.61 & 32.05 \\
        Three & 213.14 & 247.65 & 34.51 \\
        Four & 209.67 & 244.58 & 34.91 \\
        Five & 203.02 & 242.99 & 39.97 \\
        Six & 201.69 & 244.96 & 43.27 \\
        \hline
    \end{tabular}
    \end{center}
 \label{bandwidth-table-inclusion-width}
\end{table}


This characteristic allows for frequency selectivity in device design, enabling targeted operation within specific frequency bands. By combining frequency data with information on lateral and vertical confinement, the waveguide design can be tailored for specific applications. In particular, when high lateral confinement is essential to maintain coherence and controlled spin precession during acoustically mediated transport, the four-inclusion wide waveguide emerges as the optimal choice. This configuration effectively balances the need for lateral confinement and high reciprocal attenuation with a usable operational bandwidth.

\section{Geometric Parameter Optimization}

With the four-inclusion configuration established as optimal for SAW confinement, the effect of aspect ratio on waveguide performance is now considered. In this analysis, the aspect ratio of the elliptical inclusions is varied from from 3:1 to 5:1 by keeping the semi-major axis fixed at $1.7~\mum$ while adjusting the semi-minor axis from $0.566~\mum$ (for 3:1) to $0.34~\mum$ (for 5:1). This gradual change creates increasingly elongated, or ``thinner" elliptical inclusions as the aspect ratio rises.  As previously discussed, this increase in aspect ratio causes the eigenfrequencies to decrease such that at $q = 0.5$ \cite{Singh2024}, the eigenfrequencies transition from above the shear horizontal sound line to below it.  

\


\begin{figure}[tb]
    \centering
    \includegraphics[width=0.5\linewidth]{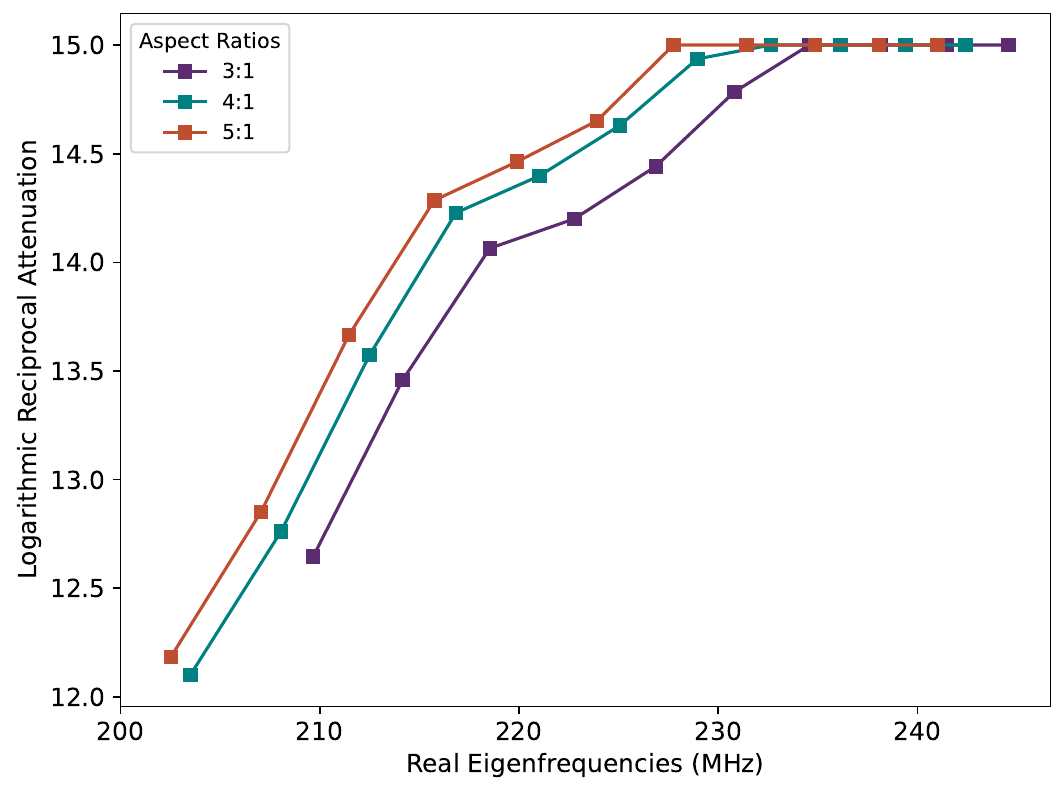}
    \caption{Operational bandwidth analysis for varying aspect ratio variation (from 3:1 to 5:1) for the four-inclusion wide waveguide. Only frequencies from eigenmodes that meet both vertical confinement (log reciprocal attenuation ${>12}$) and group velocity criteria ($v_g < \frac{v_p}{2}$) are included.}
    \label{aspect-ratio-variation}
\end{figure}


To assess the impact of aspect ratio on accessible frequencies, Figure~\ref{aspect-ratio-variation} displays the operational frequencies  for varying aspect ratios after applying the same conditions used previously: selecting values where the logarithmic reciprocal attenuation exceeds 12 and ensuring group velocity remains less than half the phase velocity. The results reveal a clear trend -- the higher aspect ratios correspond to increased bandwidth.  Table \ref{aspect-ratio-variation-table} provides a summary of the minimum and maximum eigenfrequencies for each aspect ratio and the corresponding bandwidths. The bandwidth for the 4:1 and 5:1 aspect ratios are very comparable, especially considering the resolution between data points.  The increase in bandwidth at higher aspect ratios can be attributed to mass loading \cite{Singh2024} — where added mass effectively reduces the phase velocity and, consequently, the eigenfrequency of the propagating SAW. As we make the elliptical inclusions narrower, more mass is effectively added to the surface discontinuity while reducing the overall volume of each inclusion. This reduction in volume is more pronounced when the shape is wider. With more mass added, the available volume for further reduction decreases, thereby diminishing the impact of mass loading on the surface discontinuity, which explains the limited bandwidth differential between aspect ratios of 4:1 and 5:1.  Given the small change in reciprocal attenuation between 4:1 and 5:1, the 4:1 aspect ratio may be preferred as it presents improved performance metrics over the 3:1 aspect ratio while having same bandwidth as 5:1. Moreover, fabricating the inclusion using reactive ion etching is simpler with a 4:1 aspect ratio than with a 5:1.  

\


\begin{table}[bt]
 \caption{Operational Bandwidth for Waveguides With a Varying Aspect Ratios}
    \begin{center}
    \begin{tabular}{|c|c|c|c|}
        \hline
        \textbf{Aspect Ratio} & \textbf{Minimum Frequency (MHz)} & \textbf{Maximum Frequency (MHz)} & \textbf{Bandwidth (MHz)} \\
        \hline
        3:1 & 209.67 & 244.58 & 34.91 \\
        4:1 & 203.51 & 242.42 & 38.91 \\
        5:1 & 202.53 & 241.02 & 38.49 \\
        \hline
    \end{tabular}
    \end{center}
 \label{aspect-ratio-variation-table}
\end{table}

\begin{figure}[bt]
    \centering
    \includegraphics[width=0.5\linewidth]{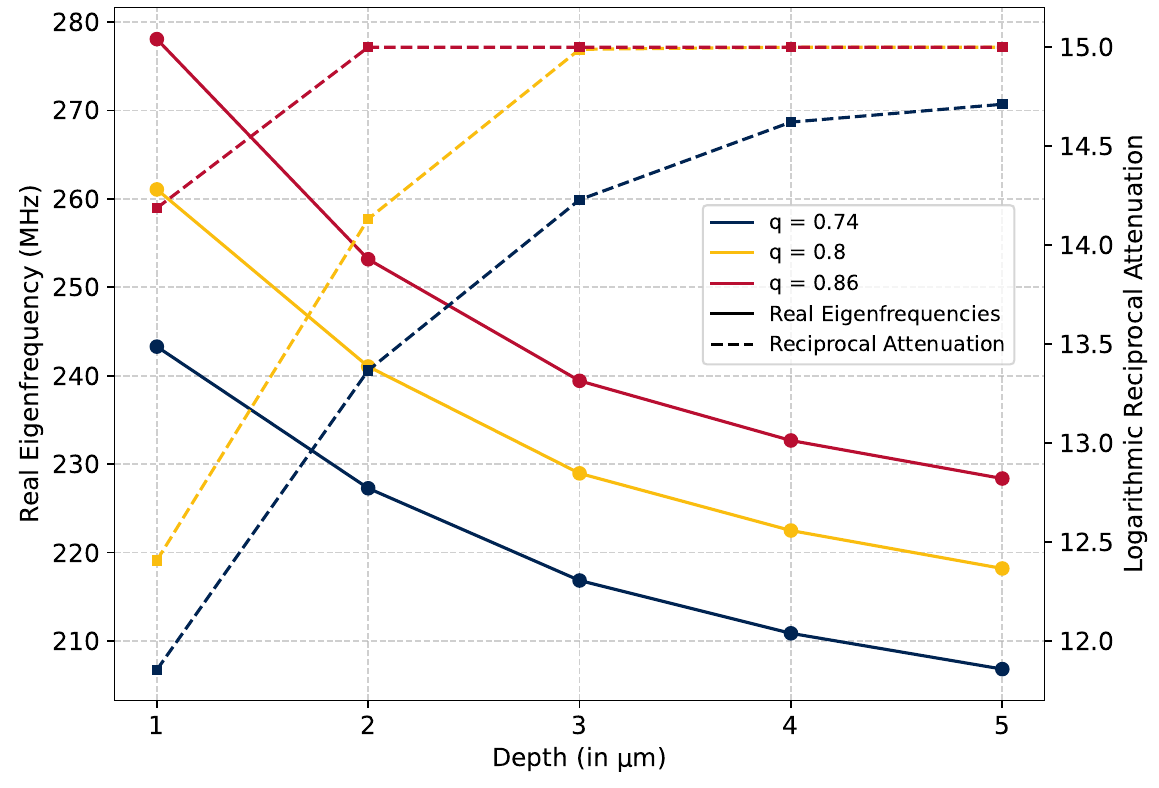}
    \caption{Variation in eigenfrequencies and logarithmic reciprocal attenuation with inclusion depth ($1~\mum$ to $5~\mum$) for the 4:1 aspect ratio elliptical cylinder inclusions at three representative reduced wavevectors ($q = 0.74,\ 0.8,\ 0.86$),}
    \label{depth-variation}
\end{figure}

To further optimize the waveguide design, the influence of inclusion depth is investigated. Figure~\ref{depth-variation} shows the relationship between inclusion depth (ranging from $1~\mum$ to $5~\mum$) for the 4:1 aspect ratio elliptical cylinder inclusions and eigenfrequency for three representative reduced wavevectors: $q$ = 0.74, 0.8, and 0.86. These $q$ values are chosen as they typically lie within the operational bandwidth of the waveguide.  Across all calculated $q$ values, a consistent trend emerges: increasing inclusion depth leads to a decrease in eigenfrequency, and an increase in logarithmic reciprocal attenuation. Surface discontinuities, such as the etched elliptical inclusions, perturb the stress-free boundary condition at the surface, thus creating local resonators and resulting in a reduction in SAW phase velocity. This effect of the inclusions is amplified at shorter wavelengths because the SAW energy is more concentrated near the surface, leading to extensive interaction with the inclusion. As the inclusion depth increases, this perturbation becomes more pronounced, leading to a further decrease in phase velocity and, consequently, a lower eigenfrequency for a given wavelength (or $q$ value).  Naturally, as the perturbation becomes more shallow, the eigenvalues increase and tend towards that of the Rayleigh mode on a bare substrate. 

\


For the $1~\mum$ inclusion depth, the eigenfrequencies are approaching the shear horizontal limit resulting in an increase of radiative coupling to bulk modes for $q=0.74$ and $q=0.8$.  There is a marked increase in the reciprocal attenuation towards the $2~\mum$, and this suggests that an inclusion depth of at least $2~\mum$ should be targeted.  While deeper inclusions further reduce the SAW eigenfrequency, fabrication limitations may constrain the achievable depth. To fabricate the inclusions in GaAs, recent developments in inductively coupled plasma/reactive-ion etching (ICP-RIE) techniques have shown that vertical-to-horizontal aspect ratios up to nine along with smooth sidewalls can be accomplished.\cite{Booker2019}  This significantly improves on previous studies with more restrictive aspect ratios and non-uniform sidewalls, where the former leads to shallower inclusions and the latter of which could introduce scattering losses and degrade waveguide performance. Further, it is critical that the inclusion not penetrate the buried semiconductor nanostructure and create edge states, which would inhibit both charge and spin transport.  Increasing the inclusion depth therefore increases the required thickness of the capping layer to avoid interactions with the nanostructure.  Therefore, to balance performance with fabrication feasibility, we propose an inclusion depth between $2~\mum$ to $3~\mum$ for the optimized waveguide design.

\section{Conclusion}

This research demonstrates a systematic approach to designing optimized PnC waveguides for enhanced SAW confinement, a critical requirement for efficient spin transport in quantum devices. Motivated by the need for precise control over propagating spins, this study addressed the challenge of minimizing energy leakage into the bulk substrate and achieving strong lateral confinement of the SAW within the waveguide core. Through finite element simulations and eigenfrequency analysis, the influence of key geometric parameters are studied, including the number of inclusions defining the waveguide width, their aspect ratio, and their depth. This analysis identified an optimal configuration – a four-inclusion wide waveguide with a $4:1$ aspect ratio and 2 to $3~\mum$ inclusion depth, which strikes a balance between maximizing both vertical and lateral confinement as well as the bandwidth of the waveguide, and ensuring efficient SAW guiding while remaining within fabrication limits. This design not only confines approximately $90\%$ of the SAW energy within the waveguide core, as quantified by the core strain energy ratio, but also minimizes lateral spread, as demonstrated by the lowest full width at half maximum of the displacement profile among the considered configurations.

\ 

Building on these findings and the concept of flying spin control gates where SAW strain acts as a contactless, tunable gate to control spin precession, this optimized waveguide design establishes a robust platform for manipulating spin information during acoustic transport.  While the length scales displayed here may be too large to achieve the desired confinement parameters for spin transport, it should be noted that the system is modeled as a linear system, and the dimensions can be reduced resulting in a corresponding increase in the frequency domain. This PnC waveguide design, coupled with continued advances in the coherent transport and manipulation of electron spins, may pave the way for the development of integrated, on-chip spin-based quantum information processors mediated by surface acoustic waves \cite{Barnes00}.


\section{Computational Modelling}

Simulations were performed using COMSOL Multiphysics, a finite element analysis (FEA) tool, to model SAW propagation in GaAs-based phononic crystal (PnC) structures. The primary analysis was carried out using eigenfrequency analysis, which determines the natural frequencies (eigenfrequencies) and corresponding mode shapes (eigenmodes) of the system. Using this approach, variations in the PnC geometry and the corresponding affect on SAW confinement and attenuation characteristics is explored. Eigenfrequencies with both real and imaginary parts were calculated, where the real part represents the oscillation frequency, and the imaginary part accounts for energy loss radiating to the bulk.

\ 

The wavevector $k$ in the dispersion relations was normalized as $q = k / k_x$, where $k_x = \pi / a$ and is called the reduced wavevector. Here, $a$ represents the lattice parameter and is set to $4~\mum$, and this normalization facilitates comparison across different lattice configurations. For all simulations, wave propagation occurred in the $\Gamma-X$ direction of the PnC, which is aligned with the [110]-direction of the GaAs crystal substrate. This was chosen as the preferred propagation direction as it is the piezoelectric direction of a GaAs (001) surface. The elasticity matrix was transformed to the [110] propagation direction using the Bond matrix rotation approach \cite{NorouzianTurner} to accurately represent anisotropic wave propagation characteristics in GaAs. Stiffness constants $c_{11}$, $c_{12}$, and $c_{44}$ were taken from Tanaka and Tamura \cite{Tanaka98} and density $\rho$ was taken from COMSOL material library. These parameter values are summarized in Table \ref{material-constants}.

\


\begin{table}[htbp]
    \centering
    \caption{Material Constants for GaAs}
    \begin{tabular}{|c|c|}
        \hline
        \textbf{Material Constant} & \textbf{Value} \\
        \hline
        Stiffness Constant, $c_{11}$ & $1.19 \times 10^{11} \, \text{Pa}$ \\
        Stiffness Constant, $c_{12}$ & $5.38 \times 10^{10} \, \text{Pa}$ \\
        Stiffness Constant, $c_{44}$ & $5.94 \times 10^{10} \, \text{Pa}$ \\
        Density, $\rho$ & $5307 \, \text{kg/m}^3$ \\
        \hline
    \end{tabular}
    \label{material-constants}
\end{table}

\begin{figure}[tb]
    \centering
    \includegraphics[width=\linewidth]{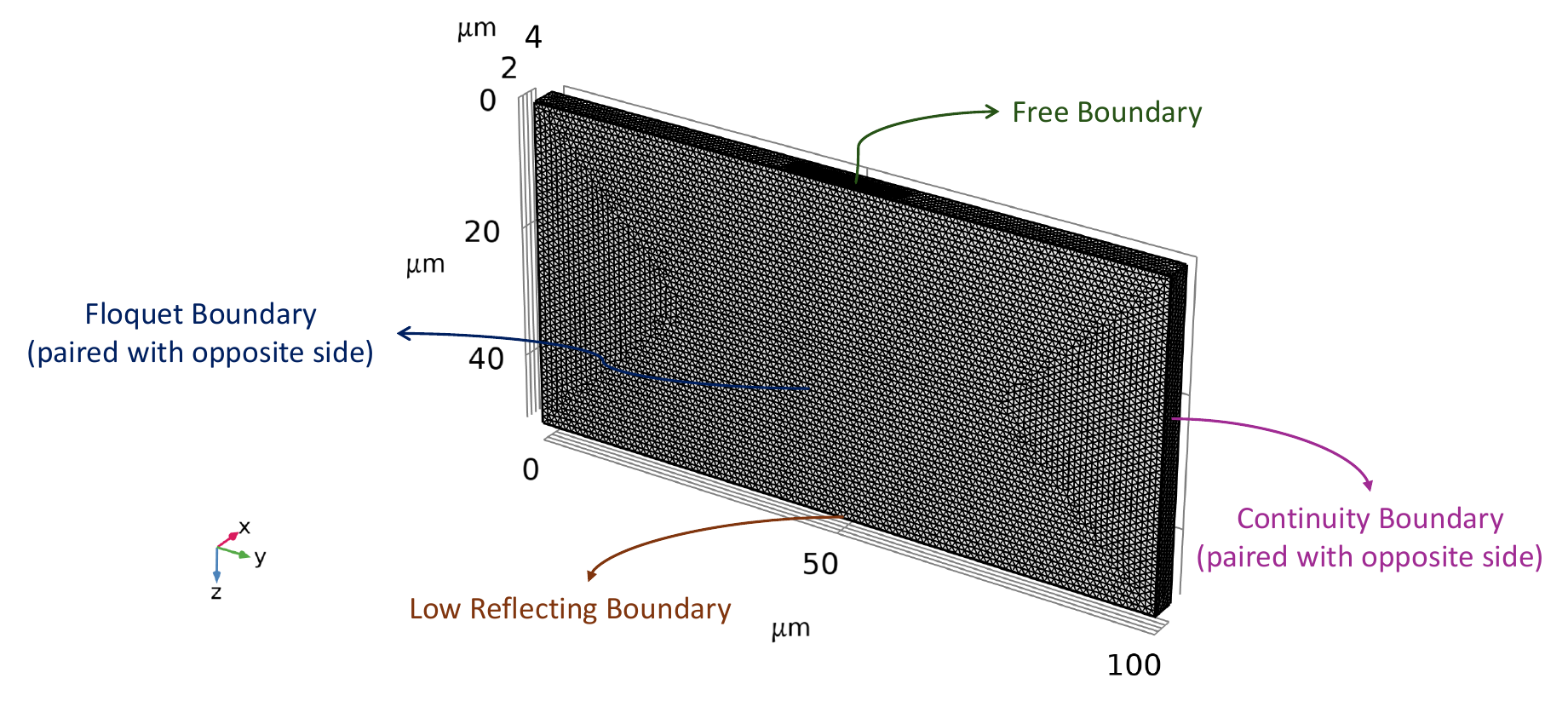}
    \caption{Mesh and boundary conditions applied in the computational domain for SAW propagation in GaAs PnCs. The domain dimensions are $4~\mum \times 100~\mum \times 50~\mum$, with mesh element sizes ranging from $1~\mum$ to $0.1~\mum$. Boundary conditions include Bloch-Floquet periodicity along the \textit{x}-axis, a Continuity boundary in the perpendicular \textit{y}-direction, a free boundary on the top surface, and a low-reflecting boundary at the bottom.}
    \label{mesh-boundary-conditions}
\end{figure}

The computational domain for the waveguide analysis consisted of a rectangular block with dimensions $4~\mum$ in the $x$-direction, $100~\mum$ in the $y$-direction, and $50~\mum$ in the $z$-direction. Inclusions were positioned in the middle of the domain, representing the PnC waveguide structure. Boundary conditions were applied as follows: Bloch-Floquet periodicity in the $x$-direction, enabling the simulation of an infinite periodic array, Continuity in the $y$-direction perpendicular to direction of wave propagation as edge effects are minimal, a Free boundary on the top surface and inclusion surfaces to allow unrestricted SAW propagation, and a Low Reflecting Boundary at the bottom to simulate a semi-infinite substrate. These boundary conditions, optimized through boundary sensitivity analysis, ensure that the simulation accurately represents physical wave confinement and minimizes reflections. Figure \ref{mesh-boundary-conditions} shows the full domain, mesh structure, and applied boundary conditions.

\


The mesh was defined based on a mesh sensitivity analysis, balancing computational efficiency and result accuracy. The maximum mesh element size was set to $1~\mum$ (corresponding to $a/4$), and the minimum element size was $0.1~\mum$ (corresponding to $a/40$), allowing for fine discretization of the domain, particularly around the inclusions. This mesh configuration ensures that even the smallest elements are sufficiently resolved, leading to accurate eigenfrequency results without excessive computational costs.

To identify relevant SAW modes, three filtering criteria were applied: the strain energy ratio, the squared polarization ratio, and the logarithmic reciprocal attenuation. The strain energy ratio measures the proportion of elastic energy confined to the top $25~\mum$ of the domain, indicating surface confinement. The squared polarization ratio quantifies the energy polarized in the \textit{xz}-plane (sagittal plane), with values above 0.6 indicating predominantly Rayleigh-like modes. Finally, logarithmic reciprocal attenuation, defined as $-\log_{10}(\mathrm{Im}(\omega) / \mathrm{Re}(\omega))$ corroborates strain energy ratio and provides insight into the radiative energy loss characteristics, with higher values indicating lower energy loss to the bulk. This also acts as our figure of merit, providing a key metric for evaluating the effectiveness of confining the eigenmode to the surface during of wave propagation across the variety of different conditions that were studied.  Taken together, these metrics ensure the selection of SAW modes that are well-confined, appropriately polarized, and exhibit minimal acoustic loss to the bulk.


\medskip
\textbf{Acknowledgements} \par 

The authors would like to gratefully recognize CMC Microsystems for the provision of products and services that facilitated this research, which includes the use of the COMSOL Design Tool.

\medskip

%
\bibliographystyle{MSP}


\end{document}